# Crystal Growth of Chalcogenides and Oxy-Chalcogenides Using Chloride Exchange Reaction

Shantanu Singh[1,2†], Boyang Zhao[1†], Christopher E. Stevens[3,4], Mythili Surendran[1], Tzu-Chi Huang[5], Bi-Hsuan Lin[5], Joshua R. Hendrickson[4], and Jayakanth Ravichandran[1,2,6*]

[1]Mork Family Department of Chemical Engineering and Materials Science, University of Southern California, Los Angeles, California, 90089, USA

[2]Core Center of Excellence in Nano Imaging, University of Southern California, Los Angeles, California, 90089, USA

[3]KBR Inc., 3725 Pentagon Blvd, Suite 110, Beavercreek, Ohio 45431, United States

[4] Sensors Directorate, Air Force Research Laboratory, Wright-Patterson Air Force Base, Ohio 45433, USA

[5]National Synchrotron Radiation Research Center, Hsinchu, 30092, Taiwan.

[6]Ming Hsieh Department of Electrical Engineering, University of Southern California, Los Angeles, California, 90089, USA

[†] These authors contributed equally: Shantanu Singh, Boyang Zhao.

[*] j.ravichandran@usc.edu

ABSTRACT

Chalcogenides and oxy-chalcogenides, including complex chalcogenides and transition metal dichalcogenides, are emerging semiconductors with direct or indirect band gaps within the visible spectrum. These materials are being explored for various photonic and electronic applications, such as photodetectors, photovoltaics, and phase-change electronics. Understanding the fundamental properties of these materials is crucial for optimizing their functionalities. Therefore, the availability of large, high-quality single crystals of chalcogenides and oxy-chalcogenides is essential for a better comprehension of their structure and properties. In this study, we present a novel crystal growth method that utilizes the exchange reaction between BaS and $ZrCl_4$/ $HfCl_4$. By carefully controlling the stoichiometric ratio of the binary sulfide to the chloride, we can grow single crystals of several materials, such as $ZrS_2$, $HfS_2$, $BaZrS_3$, and ZrOS. This method results in large single crystals with a short reaction time of 24 to 48 hours. High-resolution thin film diffraction and single-crystal X-ray diffraction confirm the quality of the crystals produced through this exchange reaction. We also report the optical properties of these materials investigated using photoluminescence and Raman measurements. The chloride exchange reaction method paves the way for the synthesis of single crystals of chalcogenides and oxy-chalcogenide systems with a short reaction time but with low mosaicity and can be an alternative growth technique for single crystals of materials that are difficult to synthesize using conventional growth techniques.

INTRODUCTION

Group IVB transition metal-based chalcogenides, including complex chalcogenides and transition metal dichalcogenides (TMDCs), have emerged as promising semiconductor materials with diverse functionalities. Zr-based chalcogenides, such as $BaZrS_3$ and $Ba_3Zr_2S_7$, are direct band gap semiconductors and excellent candidates for photovoltaic applications.[1,2] In contrast, Zr and Hf-based TMDCs, particularly $ZrS_2$ and $HfS_2$, are indirect band gap semiconductors with band gaps in the visible range, with potential for applications as channel material in field effect transistors (FETs) and photodetectors.[3,4] Additionally, ZrOS and HfOS are known to crystallize in the chiral space group $P2_13$[5,6] and are predicted to host topologically non-trivial Weyl-double phonon modes,[7] similar to FeSi, which also crystallizes in the $P2_13$ space group.[8] A crucial factor for realizing the diverse functionalities of these materials is the availability of large, high-quality single crystals, which are essential for understanding their fundamental properties. Conventional methods for single crystal growth typically involve chemical vapor transport (CVT) and halide salt-based flux growth techniques. In many instances, single-crystal growth is attained in two steps, where the polycrystalline powder of the relevant material is prepared first and then used as the precursor for crystal growth. For TMDCs like $ZrS_2$ and $HfS_2$, CVT is the most common method for bulk single-crystal growth, often utilizing $I_2$ as a transport agent.[9,10] It is important to note that while CVT produces very high-quality single crystals, the growth process can be quite slow, often taking several weeks. Single crystal growth of $BaZrS_3$ has been reported using $BaCl_2$ as a flux.[11,12] There are no reports on single crystal growth of ZrOS.

In this work, we present a novel chloride exchange reaction mechanism that yields high-quality single crystals of chalcogenide/oxy-chalcogenide materials with a short reaction time of 24 to 48 hours. An alkaline earth metal sulfide (BaS) supplies sulfur (and Ba in the case of $BaZrS_3$), while a transition metal tetrachloride (MCl₄, where M = Zr or Hf) provides the metal component. At high temperatures, these precursors undergo an exchange reaction that yields BaCl₂ and the desired metal sulfide (MS₂), eliminating the need for pre-synthesized polycrystalline MS₂. A significant advantage of this approach is the ability to tailor material stoichiometry simply by adjusting the molar ratios of the precursors. Using a 1:2 molar ratio of MCl₄ to BaS results in the formation of large single crystals of MS₂ within approximately two days. By employing a 1:1 reactant molar ratio, we grew large, mm-scale single crystals of zirconium oxysulfide (ZrOS). Furthermore, altering the ratio to 1:3 shifts the reaction toward the formation of BaZrS₃ single crystals. Comprehensive chemical, structural, and optical characterizations confirm the excellent quality of the obtained single crystals. Overall, this chloride exchange reaction method introduces a fast and efficient growth technique while providing precise stoichiometric control, paving the way for synthesizing a diverse array of complex chalcogenides and oxysulfide materials.

RESULTS AND DISCUSSIONS

**Exchange Reaction Growth of Transition Metal Dichalcogenides**

Large single crystals of transition metal dichalcogenide (MS₂, M = Zr, Hf) were obtained by an exchange reaction between barium sulfide and transition metal chloride in a 2:1 molar ratio, using a two-zone furnace. As the temperature increases, an exchange reaction happens between the reactants, resulting in the formation of MS₂ and BaCl₂, according to the following equation:

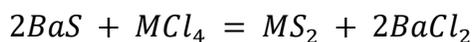

$$2BaS + MCl_4 = MS_2 + 2BaCl_2$$

Large single crystals of ZrS₂ and HfS₂ of cm-scale were obtained, both in the form of hexagon platelet-like single crystals but of different colors. ZrS₂ crystals are shiny with silver color, while HfS$_2$ are shiny with orange color, as shown in **Figure 1(a & b)**. In both cases, the reactants were kept at the high-temperature end of the quartz ampoule, and the MS$_2$ crystallized in the high-temperature zone, indicating that the formation of MS$_2$ crystals is an endothermic reaction based on Le Chatelier's principle. To confirm this, synthesis experiments for HfS$_2$ were performed by reversing the temperature profile so that the reactants were kept at a lower temperature. In this case, large HfS$_2$ crystals were obtained away from the initial reactant position at the other end of the ampoule in the high-temperature zone (see **Supporting Information, I**). The chemical composition and homogeneity of the crystals were confirmed using energy dispersive spectroscopy (EDS) in a scanning electron microscope (SEM). EDS spectra were collected for several crystals to confirm that the MS$_2$ crystals have an M: S ratio close to the expected stoichiometry (1:2). Representative EDS spectra for MS$_2$ crystals are shown in **Figure S2(a & b)**. It is important to note that Ba or Cl was not detected in the MS$_2$ crystals down to the EDS detection limits, confirming that the single crystals obtained by this are of high composition purity and no trace elements from the byproduct of the exchange reaction (BaCl$_2$) were incorporated into the crystals. EDS elemental maps were collected for both materials, indicating a uniform element distribution of transition metal and sulfur in the sample **(Figure S3)**.

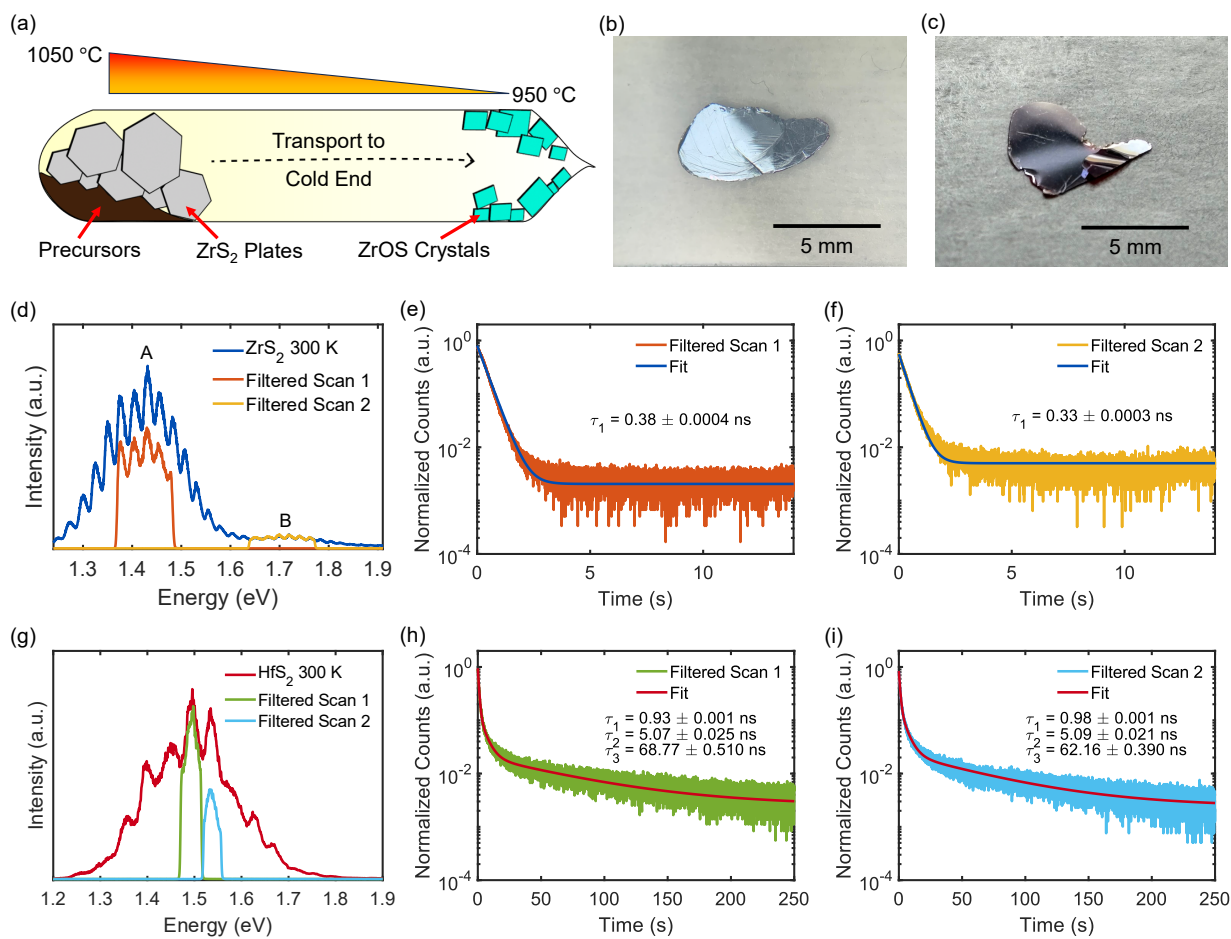

**Figure 1:** Exchange reaction synthesized transition metal dichalcogenide (TMDC) single crystals. **(a)** Schematic illustrating the crystal growth of TMDCs and ZrOS. Optical images of **(b)** $ZrS_2$, and **(c)** $HfS_2$ single crystals. Photoluminescence spectra for **(d)** $ZrS_2$ and **(g)** $HfS_2$ collected using 470 nm laser excitation at 300 K, showing the filtered scans used for TRPL studies. TRPL measurements and fits for filtered scans for **(e, f)** $ZrS_2$ and **(h, i)** $HfS_2$, respectively.

Photoluminescence (PL) spectra for the $MS_2$ crystals were collected at room temperature (RT) using a continuous-wave laser excitation of 470 nm. In the case of $ZrS_2$, as shown in **Figure 1(d)**, two features were observed in the PL spectra, along with a periodic fluctuation of intensity as a function of energy. The periodic fluctuation component is attributed to Fabry-Pérot interference fringes, resulting from interference between the two surfaces of the crystal. The peak values were approximated using a sum of two Gaussian functions to fit the spectrum. The low-energy peak is centered around 1.43 eV, and the high-energy peak is around 1.7 eV. The 1.7 eV peak closely corresponds to the indirect band gap reported for bulk $ZrS_2$ samples in the literature.[3,9,13] Presumably, the peak at 1.43 eV could correspond to some defect level. In the case of $HfS_2$ (**Figure 1(g)**), a single PL peak is observed, centered around 1.48 eV, along with a periodic fluctuation of intensity due to the interference fringes, similar to $ZrS_2$. This value is similar to the PL peak reported in the literature for $HfS_2$[14,15] and is attributed to the indirect band gap of the material, making $HfS_2$ a good candidate for near-infrared photodetection.

To further investigate the optical properties of $HfS_2$ and $ZrS_2$, time-resolved photoluminescence (TRPL) measurements were performed using a 470nm variable repetition rate picosecond diode laser, a single photon avalanche diode, and an electronic timing box. The TRPL measurement results are shown in **Figure 1**. To the best of our knowledge, TRPL results have not been reported before for $ZrS_2$ and $HfS_2$. In the case of $ZrS_2$, TRPL measurements were performed for filtered PL scans centered around the 1.43 eV peak (peak A) and 1.7 eV peak (peak B), as shown in **Figure 1(e, f)**. At RT, a single exponential decay model was used for fitting, yielding a lifetime of 0.38 ns for peak A and a relatively shorter 0.33 ns lifetime for peak B. This distinction becomes more evident in the TRPL measurements performed at 4 K, as shown in **Figure S8 (a-c)**. Both peaks exhibit longer relaxation lifetimes at 4 K compared to RT due to carrier localization and reduced phonon interactions.[16] A bi-exponential model was used to fit the TRPL decay for both peaks. An average lifetime of ~0.86 ns is observed for peak A, while a much shorter average lifetime of ~0.36 ns is observed for peak B. The shorter lifetime of a few hundreds of picoseconds has been attributed to inter-band transition in the case of TMDCs, while the longer lifetimes have been assigned to defect-related recombination.[17–19] Thus, the shorter lifetime for peak B indicates that it corresponds to the band gap of $ZrS_2$, while peak A might correspond to a defect-mediated transition.

For HfS$_2$, as shown in **Figure 1(h, i)**, a tri-exponential decay model gave the best fit for the measured data, with the three lifetimes being approximately 0.9-1 ns, 5 ns, and 60-70 ns, respectively. However, the major (~99.9%) contribution is from the shortest lifetime (~0.9-1 ns). Thus, the short lifetime (average lifetime in the case of HfS$_2$) indicates the high quality of the exchange reaction-grown single crystals of ZrS$_2$ and HfS$_2$. For TRPL measurements at 4 K, all three lifetimes observed are longer, with the values being 1.72 ns, 8.67 ns, and 92.66 ns. More details on the fitting results can be found in the **Supporting Information**.

**Characterization of Single Crystals of Exchange Reaction-Grown TMDCs**

The structural quality of the single crystals of ZrS$_2$ and HfS$_2$ was determined by employing a combination of high-resolution thin film and single-crystal X-ray diffraction (SC-XRD) measurements. To confirm the out-of-plane orientation of the MS$_2$ crystals, high-resolution thin film diffraction measurements were performed by aligning to the presumed crystal orientation. The out-of-plane diffraction patterns for ZrS$_2$ and HfS$_2$ are shown in **Figure 2(a) and (b),** respectively, and show the family of peaks corresponding to the 00*l*-type peaks, confirming that for both MS$_2$ crystals, the *c*-axis lies out-of-plane. The rocking curve (RC) measurement for the 001 peak for ZrS$_2$ is shown in **Figure S5(a)**. An approximate full width at half maximum (FWHM) value of ~0.13° is observed. More than one reflection might be observed in the case of a twinned sample, which is possible for large, cm-scale single crystals of TMDCs. Similarly, for HfS$_2$, the RC was collected for the 002 peak. The twinning is also evident in the case of HfS$_2$ single crystal. The FWHM of the RC for the 002 reflection is ~0.06°, which is comparable to the FWHM values reported for bulk HfS$_2$ single crystals.[20]

Single crystal diffraction measurements were performed at 290 K for ZrS$_2$ and HfS$_2$ using Mo K-$\alpha$ X-ray radiation. In the case of ZrS$_2$, the crystal structure was refined to the trigonal $P\bar{3}m1$ space group, with lattice parameters being $a = b = 3.66$ Å, and $c = 5.83$ Å. These values match well with the lattice parameters reported in the literature.[21] The precession map for ZrS$_2$ extracted from the SC-XRD data along the $hk0$ planes is shown in **Figure 2(c)**. The precession maps reveal the expected hexagonal symmetry of the ZrS$_2$ crystal structure. As shown in **Figure 2(d)**, the crystal structure of ZrS$_2$ consists of layers of edge-sharing ZrS$_6$ octahedra units in the $a$-$b$ plane, with the Zr-Zr and Zr-S bond lengths being 3.66 and 2.56 Å, respectively. These layers are connected by weak van der Waals interaction along the $c$-axis, which is characteristic of two-dimensional layered materials. HfS$_2$ is isostructural to ZrS$_2$ and is also refined to $P\bar{3}m1$ space group, with lattice parameters being $a = b = 3.63$ Å, and $c = 5.85$ Å. The Hf-Hf and Hf-S bond lengths are 3.63 and 2.54 Å, respectively. More details on data collection, reduction, and refinement are given in **Supporting Information, VI**.

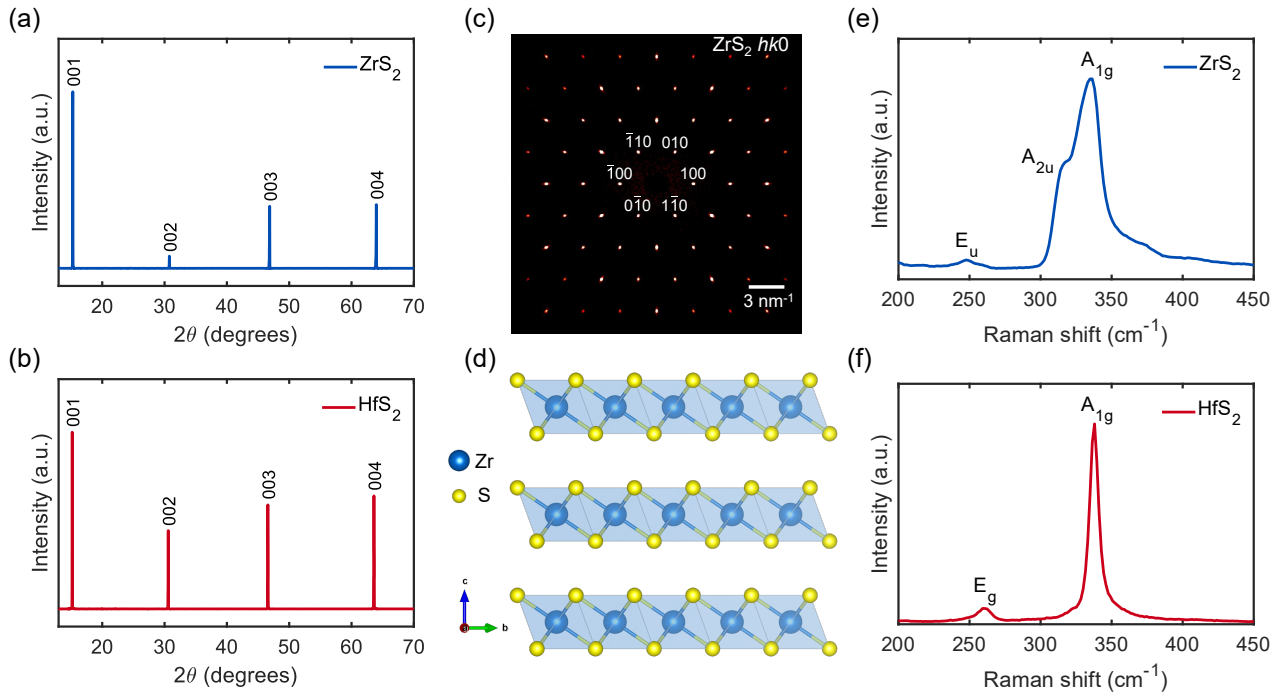

**Figure 2:** Characterization of TMDC single crystals. Out-of-plane X-ray diffraction patterns of **(a)** ZrS$_2$ and **(b)** HfS$_2$ show (00$l$) texture. **(c)** The $hk0$ precession map for ZrS$_2$ generated from SC-XRD data; the circles have been added to highlight selected peaks. **(d)** Schematic of the ZrS$_2$ crystal structure, showing the layered structure along $c$-axis as derived from single crystal diffraction. Raman spectra of **(e)** ZrS$_2$ and **(f)** HfS$_2$ collected at room temperature using a 532 nm laser excitation.

Raman spectroscopy measurements were performed at RT using a 532 nm laser excitation to study the vibrational modes of the MS$_2$ materials. A representative Raman spectrum of ZrS$_2$ single crystal is shown in **Figure 2(e)**. Three features are observed for ZrS$_2$ at 248 cm$^{-1}$, 316 cm$^{-1}$, and 335 cm$^{-1}$, which can be assigned to the $E_g$, $A_{2u}$, and $A_{1g}$ modes, respectively.[21,22] The Raman spectra for HfS$_2$ is reported in **Figure 2(f)** indicates two features at 262 cm$^{-1}$ and 338 cm$^{-1}$, which correspond to the $E_g$ and $A_{1g}$ modes, respectively.[23,24]

**Crystal Growth of Oxy-Chalcogenides**

ZrOS single crystals were grown by increasing the relative amount of ZrCl$_4$ in the reaction mixture compared to the ratio used for ZrS$_2$ growth. Specifically, a ZrCl$_4$: BaS molar ratio of 1:1 was employed. A temperature profile similar to the one used to synthesize TMDCs was followed. Large ZrOS single crystals were obtained on the cold end of the ampoule, along with some ZrS$_2$ single crystals on the hot end, as shown in **Figure 1(a)**. The oxygen in ZrOS is potentially incorporated from the reaction of chloride in the reaction mix with the quartz (SiO$_2$) ampoule used as the reaction vessel. The formation of ZrOS appears to be an exothermic process, as the crystallization is observed at the colder end of the ampoule. ZrOS single crystals grow in the form of 0.1- 1 mm blocks with shiny, flat facets (later determined from SC-XRD to be 111 and 121, as shown in **Figure S6(a)**). While thinner ZrOS crystals (~ 100 μm) look transparent to incandescent light, large pieces weakly glow blue under fluorescent light, as shown in **Figure 3(a)**. Similar attempts to synthesize HfOS single crystals were unsuccessful.

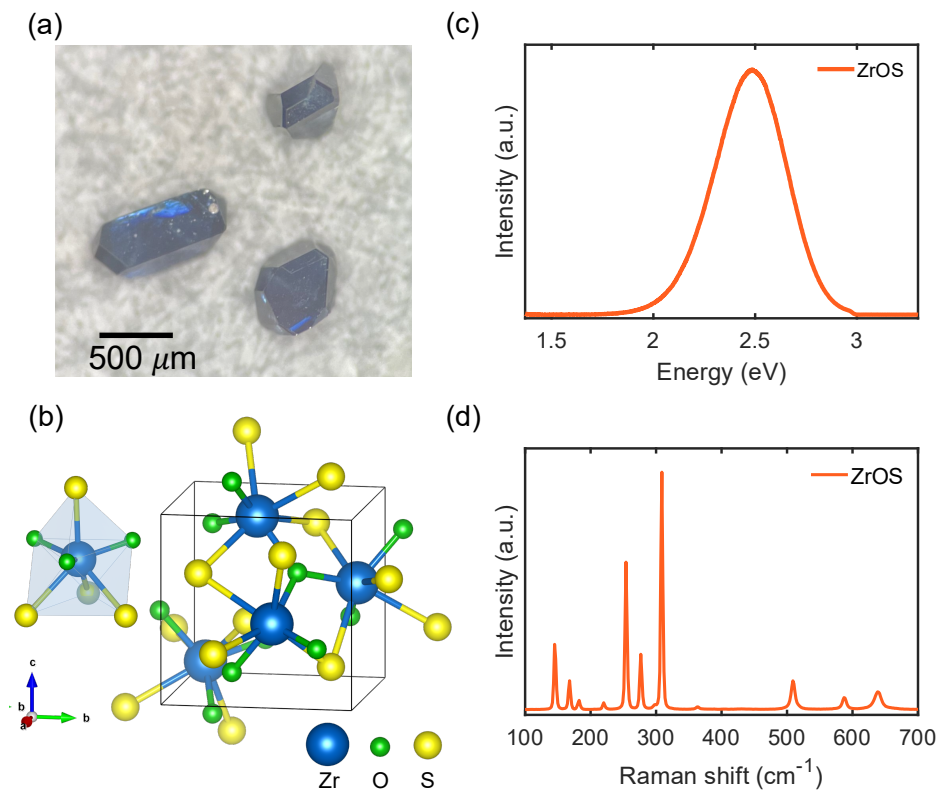

**Figure 3:** Exchange reaction synthesized ZrOS single crystals. **(a)** Optical image of ZrOS **(b)** Crystal structure visualization for ZrOS, top left shows the Zr coordination polyhedral. **(c)** PL spectrum of ZrOS measured using a 375 nm pulsed laser excitation. **(d)** Raman spectrum of ZrOS.

To the best of our knowledge, this is the first report on the synthesis of ZrOS single crystals, though powder samples have been reported before.[5,25] EDS measurements were done to confirm the elemental stoichiometry and homogeneity of the ZrOS crystals. EDS spectrum confirms a 1:1 elemental ratio of Zr: S, and mapping results show a uniform distribution of Zr and S throughout the crystal, as shown in **Figure S2(c)**. High-resolution thin film diffraction measurements were carried out to learn about the out-of-plane orientation for ZrOS crystals. As shown in **Figure S6(b)**, 111 and 222 reflections are observed, confirming that the large, flat facets in the crystals correspond to the direction of the body diagonal of the cubic crystal structure. Rocking curve measurement for the 111 reflection gives an FWHM value of 0.019°, confirming the large coherent domains in the single crystals. We performed SC-XRD measurements to analyze the crystal structure of the material. ZrOS crystallizes in cubic $P2_13$ space group, with lattice parameters being $a = b = c = 5.7$ Å. The crystal structure is shown in **Figure 3(b)**. The coordination number for the zirconium atom in the structure is 7. The resulting heptahedron around the zirconium atom is made of 3 oxygen atoms and 4 sulfur atoms. The Zr-O bond length is 2.08 Å and is equal for all three bonds. In the case of Zr-S bonds, a longer bond length of 2.72 Å is observed for three Zr-S bonds, while a shorter bond length of 2.59 Å is observed for the remaining three Zr-S bonds. As ZrOS crystallizes in a chiral Sohncke space group, it makes ZrOS an excellent material system for understanding light-matter interaction in chiral material systems.[26] Furthermore, as reported for other materials crystallizing in the $P2_13$ space group, ZrOS is a potential candidate for hosting unconventional chiral quasiparticles such as double-Weyl phonons.[7]

PL measurements were performed using a 375 nm pulsed laser excitation on ZrOS, and a clean peak was observed around 2.48 eV, as shown in **Figure 3(c)**, potentially causing the blue glow observed in ZrOS crystals. This observation qualitatively agrees with the notion that the band gap of ZrOS would be expected to lie in between $ZrS_2$ (> 2 eV) and $ZrO_2$ (< 3eV).[27,28] This was also substantiated with X-ray excited optical luminescence (XEOL) using focused synchrotron X-rays of 9.67 keV as excitation radiation. No additional PL peaks were observed at higher energies, as shown in **Figure S9**. Raman measurements were performed using a 532 nm laser excitation at RT. Signature peaks are observed in the 80 to 700 cm$^{-1}$ range, at 145 cm$^{-1}$, 168 cm$^{-1}$, 183 cm$^{-1}$, 220 cm$^{-1}$, 254 cm$^{-1}$, 278 cm$^{-1}$, 309 cm$^{-1}$, 363 cm$^{-1}$, 509 cm$^{-1}$, 589 cm$^{-1}$, and 640 cm$^{-1}$.

**Crystal Growth of Perovskite Chalcogenides**

To achieve single crystal growth of the chalcogenide perovskite BaZrS$_3$, barium sulfide and zirconium chloride were used as precursors in a 3:1 molar ratio. The exchange reaction between BaS and ZrCl$_4$ results in the formation of BaCl$_2$, which also acts as a flux for the growth of single crystals of BaZrS$_3$ according to the following reaction:

$$3BaS + ZrCl_4 = BaZrS_3 + 2BaCl_2$$

Large single crystals of BaZrS$_3$ of ~100 – 400 $\mu$m in size with flat surfaces are obtained. The SEM image of a representative crystal of BaZrS$_3$ is shown in **Figure 4(a)**. EDS spectrum indicates an elemental ratio of 1:1:3 for Ba: Zr: S in these crystals, while elemental mapping confirms a uniform distribution for Ba, Zr, and S in the crystals. High-resolution thin film diffraction scan reveals 040-type reflections, indicating 0k0, where k is even, to be the out-of-plane orientation for the BaZrS$_3$ single crystals, as shown in **Figure S7**. Rocking curve measurement for the 040 reflection gives an FWHM value of 0.016°, which is comparable to the value reported in literature.[11] This confirms the large coherent domain sizes of the BaZrS$_3$ crystals obtained using the exchange reaction method. The crystal structure for BaZrS$_3$ was determined using SC-XRD measurements. BaZrS$_3$ crystallizes in the orthorhombic *Pnma* space group in a distorted perovskite structure, with lattice parameters $a$ = 7.06 Å, $b$ = 9.98 Å, and $c$ = 7.02 Å. As observed in the crystal structure of BaZrS$_3$ shown in **Figure 4(b)**, each Zr atom is bonded to 6 S atoms, resulting in an octahedral coordination for Zr. The details of structure refinement results are given in **Supporting Information.**

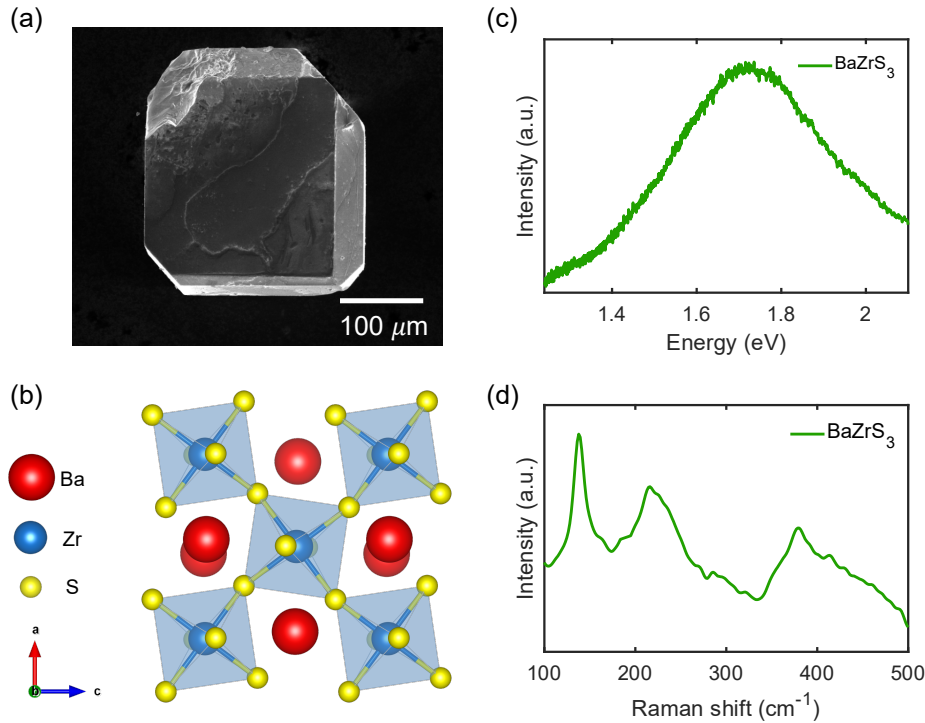

**Figure 4:** BaZrS$_3$ single crystals obtained through exchange reaction. **(a)** SEM image of BaZrS$_3$ crystal. **(b)** Visualization of crystal structure of BaZrS$_3$ with *b*-axis out of plane. **(c)** PL and **(d)** Raman spectrum of BaZrS$_3$ collected at room temperature using 532 nm laser excitation.

The PL spectrum for BaZrS$_3$ crystal is shown in **Figure 4(c)** and indicates a single broad peak centered around ~1.73 eV, with an FWHM of ~450 meV. The band gap value is similar to the values reported for crystals and thin films of BaZrS$_3$.[11,29] The Raman spectrum for BaZrS$_3$ indicates several peaks, as shown in **Figure 4(d)**. The sharp peak around 138 cm$^{-1}$ can be assigned to the $A_g^4$ mode, the broad peak around 215 cm$^{-1}$ can be majorly attributed to a combination of $B_{2g}^6$ and $A_g^6$ modes, while the broad peak around 380 cm$^{-1}$ can be majorly attributed to a combination of $B_{1g}^5$ and $B_{2g}^7$ modes.[30,31]

CONCLUSIONS

We have successfully synthesized single crystals of several binary and ternary transition metal chalcogenides and oxy-chalcogenides using a novel exchange reaction method that employs barium sulfide and transition metal tetrachloride as precursors. By carefully controlling the molar ratios of the reactants, we achieved different compositions, including transition metal dichalcogenides (TMDCs) such as $ZrS_2$ and $HfS_2$, an oxysulfide (ZrOS), and $BaZrS_3$. This method has enabled us to grow large single crystals with large coherent domains but in a short reaction time of 24 to 48 hours. The quality of the crystals obtained has been confirmed by various chemical, structural, and optical measurements. This synthesis route offers a unique pathway for the growth of complex chalcogenide/ oxy-chalcogenide compositions. By utilizing binary sulfide and chloride as reactants, we can achieve high-quality single crystals with varying stoichiometries by controlling their ratios.

METHODS

***Exchange Reaction Crystal Growth:*** Single crystal samples were synthesized by utilizing an exchange reaction between barium sulfide powder (Sigma-Aldrich, 99.9%) and zirconium chloride powder or hafnium chloride powder, which were stored and handled in a nitrogen-filled glove box. For $ZrS_2$/$HfS_2$, stoichiometric quantities (2:1) of precursor powders (BaS: $ZrCl_4$/ $HfCl_4$) with a total weight of 1.0 g were mixed and loaded into a quartz ampoule of an inner diameter of 15 mm and thickness 2 mm inside the glove box. A BaS: $ZrCl_4$ stoichiometry of 3:1 was used for $BaZrS_3$, while a 1:1 ratio of precursors was used for ZrOS synthesis. The quartz ampoule was capped with ultra-torr fittings and a bonnet needle valve to avoid exposure to the air. The ampoule was then evacuated and sealed using a blowtorch, with oxygen and natural gas as the combustion mixture. The sealed ampoules were loaded and heated in a Lindberg/Blue M Mini-Mite Tube Furnace. During the synthesis, precursors were heated to the reaction temperature of 1050 °C (cold end at 950 °C) at 100 °C/h and held for 24 – 48 h before cooled down back to RT at 100 °C/h.

***Structural Characterization:*** The high-resolution out-of-plane XRD scans were collected in a Bruker D8 Advance X-ray diffractometer with Cu Kα radiation with a power setting of 40 kV and 40 mA using a Ge (004) two bounce monochromator at RT in the thin film out-of-plane configuration. Crystals were loaded on the Compact Cradle on top of a glass-slide holder. Rocking curve scans were collected by varying the incident angle while keeping the 2θ fixed at the center of the peak position.

Single crystal XRD measurements were done using a Rigaku XtaLAB Synergy-S diffractometer equipped with Dual source (Cu and Mo, PhotonJet-S) and a HyPix-6000HE detector, using Mo Kα radiation (0.7107 Å) with a power setting of 50 kV and 1 mA. The diffractometer was equipped with an Oxford Cryosystems Cryostream 800, which was used for low-temperature measurements. Crystals were mounted on MiTeGen Kapton loops (Dual Thickness MicroMounts$^{TM}$) and placed on the goniometer head. Data reduction, scaling, and precession map analysis were done in CrysAlisPro, and crystal structures were solved and refined in ShelXle.

***Optical and Chemical Characterization:*** Secondary electron imaging and energy dispersive X-ray spectroscopy (EDS) measurements were done in a Nova NanoSEM 450 Field Emission Scanning Electron Microscope equipped with an OXFORD Ultim Max detector. Single crystals of the relevant materials were mounted onto conductive silicon pieces using carbon tape. EDS data was collected using an acceleration voltage of 20kV, and data analysis was done using the Aztec software.

The RT PL spectroscopy and Raman measurements were performed in a Renishaw inVia confocal Raman microscope using a 532 nm diode laser through a 50× objective with a numerical aperture of 0.1. The excitation power used was ~1 mW. The Ultraviolet photoluminescence (UV-PL) and X-ray excited optical luminescence (XEOL) experiments were conducted at the X-ray Nanoprobe 23A beamline of the Taiwan Photon Source (TPS) at National Synchrotron Radiation Research

Center (NSRRC), Taiwan. XEOL spectra were collected with multi-mode optical fiber (400 μm core diameter) attached to a HORIBA iHR 550 spectrometer with deep thermoelectric cooled CCD (Syncerity BI UV-Vis) with a resolution of 2048 ×70 pixels. A picosecond pulsed laser (PLP-10 Hamamatsu) with emission wavelength at 375 nm was used as the PL excitation source. The emitted PL light was collected with a 2 m optical fiber and directed to the spectrometer (iHR320, Horiba). The PL spectrum was recorded with a charge-coupled device (Syncerity BI UV-Vis) with a resolution of 2048 × 70 pixels.

RT and 4K PL and time-resolved photoluminescence (TRPL) measurements were performed by loading the samples into a Montana S200 cryostation with a 0.9 NA *in-situ* objective. The samples were excited by either a tunable ti:sapphire fs laser (Coherent Chameleon with OPO and SHG attachment) or a 470nm diode laser with tunable repetition rate (PicoQuant, LDH-IB-470-B). The resulting photoluminescence was sent to a grating spectrometer (Princeton Instruments, SP-2750) and measured via a CCD (PyLoN 400BRX). For the TRPL measurements, the PL was filtered by tunable long and short pass Semrock filters. The filtered light was sent to a single photon avalanche detector and HydraHarp timing box (Picoquant), resulting in the timing characteristics of the emission.

ASSOCIATED CONTENT

**Supporting Information**.

Temperature control for crystal growth, EDS Measurements, High-Resolution Thin Film Diffraction Measurements, TRPL Measurements on $ZrS_2$ and $HfS_2$, ZrOS XEOL measurements, SC-XRD Measurement and Refinement Details (PDF)

AUTHOR INFORMATION

**Corresponding Author**

Jayakanth Ravichandran

Mork Family Department of Chemical Engineering and Materials Science, University of Southern California, Los Angeles, California 90089, United States; Ming Hsieh Department of Electrical and Computer Engineering, University of Southern California, Los Angeles, California 90089, United States; Core Center of Excellence in Nano Imaging, University of Southern California, Los Angeles, California 90089, United States

E-mail: j.ravichandran@usc.edu

**Authors**

Shantanu Singh

Mork Family Department of Chemical Engineering and Materials Science, University of Southern California, Los Angeles, California 90089, United States; Core Center of Excellence in Nano Imaging, University of Southern California, Los Angeles, California 90089, United States

Boyang Zhao

Mork Family Department of Chemical Engineering and Materials Science, University of Southern California, Los Angeles, California 90089, United States

Christopher E. Stevens

KBR Inc., 3725 Pentagon Blvd, Suite 210, Beavercreek, Ohio 45431, United States; Air Force Research Laboratory, Sensors Directorate, Wright-Patterson Air Force Base, 2241 Avionics Circle, Ohio 45433, United States

Mythili Surendran

Mork Family Department of Chemical Engineering and Materials Science, University of Southern California, Los Angeles, California 90089, United States

Tzu-Chi Huang

National Synchrotron Radiation Research Center, Hsinchu, 30092, Taiwan

Bi-Hsuan Lin

National Synchrotron Radiation Research Center, Hsinchu, 30092, Taiwan

Joshua R. Hendrickson

Air Force Research Laboratory, Sensors Directorate, Wright-Patterson Air Force Base, 2241 Avionics Circle, Ohio 45433, United States

**Author Contributions**

S.S., B.Z., and J.R. conceived the idea and designed the experiments. S.S. and B.Z. synthesized the single crystals and performed the structural characterization measurements. S.S. and M.S. performed room temperature PL and Raman measurements. C.E.S. and J.R.H. performed PL and TRPL measurements. T.C.H. and B.H.L. performed pulsed laser PL and XEOL measurements. All authors discussed the results. S.S. and J.R. wrote the manuscript with input from all other authors.

**Notes**

The authors declare no competing financial interest.

ACKNOWLEDGMENT

This work, in part, was supported by the Army Research Office (ARO) MURI program with award number W911NF-21-1-0327 and an NSF grant with award number DMR-2122071. The crystal growth and characterization tools in part were supported by an ONR grant with award number N00014-23-1-2818.

The authors gratefully acknowledge the use of facilities at Dr. Stephen Cronin's Lab and Core Center for Excellence in Nano Imaging at the University of Southern California for the results reported in this manuscript. An NSF grant with award number CHE-2018740 provided funds to acquire the Rigaku XtaLAB Synergy-S diffractometer that was used for single-crystal XRD studies.

J.R.H. acknowledges support from the Air Force Office of Scientific Research under award number FA9550-25RYCOR006. The research performed by C.E.S. at the Air Force Research Laboratory was supported by contract award FA807518D0015.

TOC Graphic

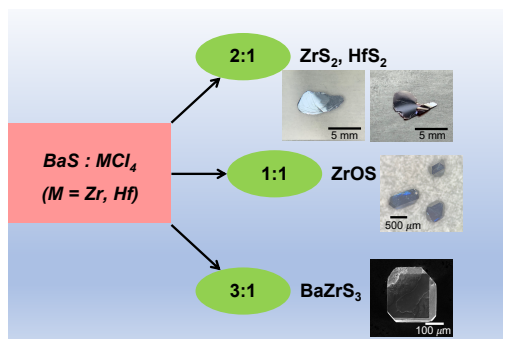

# SUPPORTING INFORMATION

# Crystal Growth of Chalcogenides and Oxy-Chalcogenides Using Chloride Exchange Reaction

Shantanu Singh[1,2], Boyang Zhao[1], Christopher E. Stevens[3,4], Mythili Surendran[1], Tzu-Chi Huang[5], Bi-Hsuan Lin[5], Joshua R. Hendrickson[4], and Jayakanth Ravichandran[1,2,6*]

[1]Mork Family Department of Chemical Engineering and Materials Science, University of Southern California, Los Angeles, California, 90089, USA
[2]Core Center of Excellence in Nano Imaging, University of Southern California, Los Angeles, California, 90089, USA
[3]KBR Inc., 3725 Pentagon Blvd, Suite 110, Beavercreek, Ohio 45431, United States
[4] Sensors Directorate, Air Force Research Laboratory, Wright-Patterson Air Force Base, Ohio 45433, USA
[5]National Synchrotron Radiation Research Center, Hsinchu, 30092, Taiwan.
[6]Ming Hsieh Department of Electrical Engineering, University of Southern California, Los Angeles, California, 90089, USA

CONTENTS

## I. Temperature Control for Endothermic Crystal Formation for TMDCs

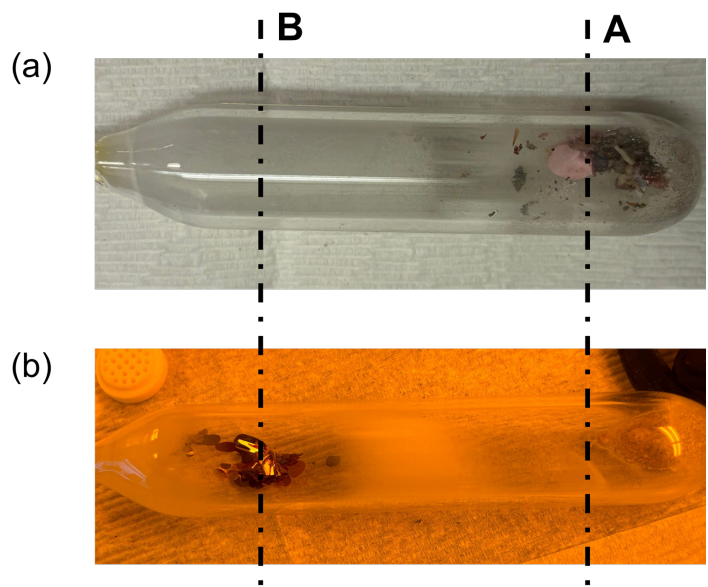

**Figure S1:** HfS$_2$ crystal growth with reversed temperature profile. **(a)** 1050 °C maintained at A zone and 950 °C at B zone, and **(b)** 1050 °C maintained at B zone and 950 °C at A zone.

To confirm the endothermic nature of the formation of single crystals of TMDCs, synthesis experiments were done with reversed temperature profile for HfS$_2$. As the temperature profile is reversed, the single crystal formation zone changes from A to B, i.e., the single crystals are formed in the high-temperature zone, confirming the endothermic nature of crystallization. Furthermore, this control of the growth zone by controlling the temperature profile for vapor transport allows the growth of clean single crystals of TMDCs, away from the BaCl$_2$. This ensures that the crystals obtained do not need to be washed/ cleaned to get rid of any unwanted impurities.

## II. EDS Measurements on Single Crystals

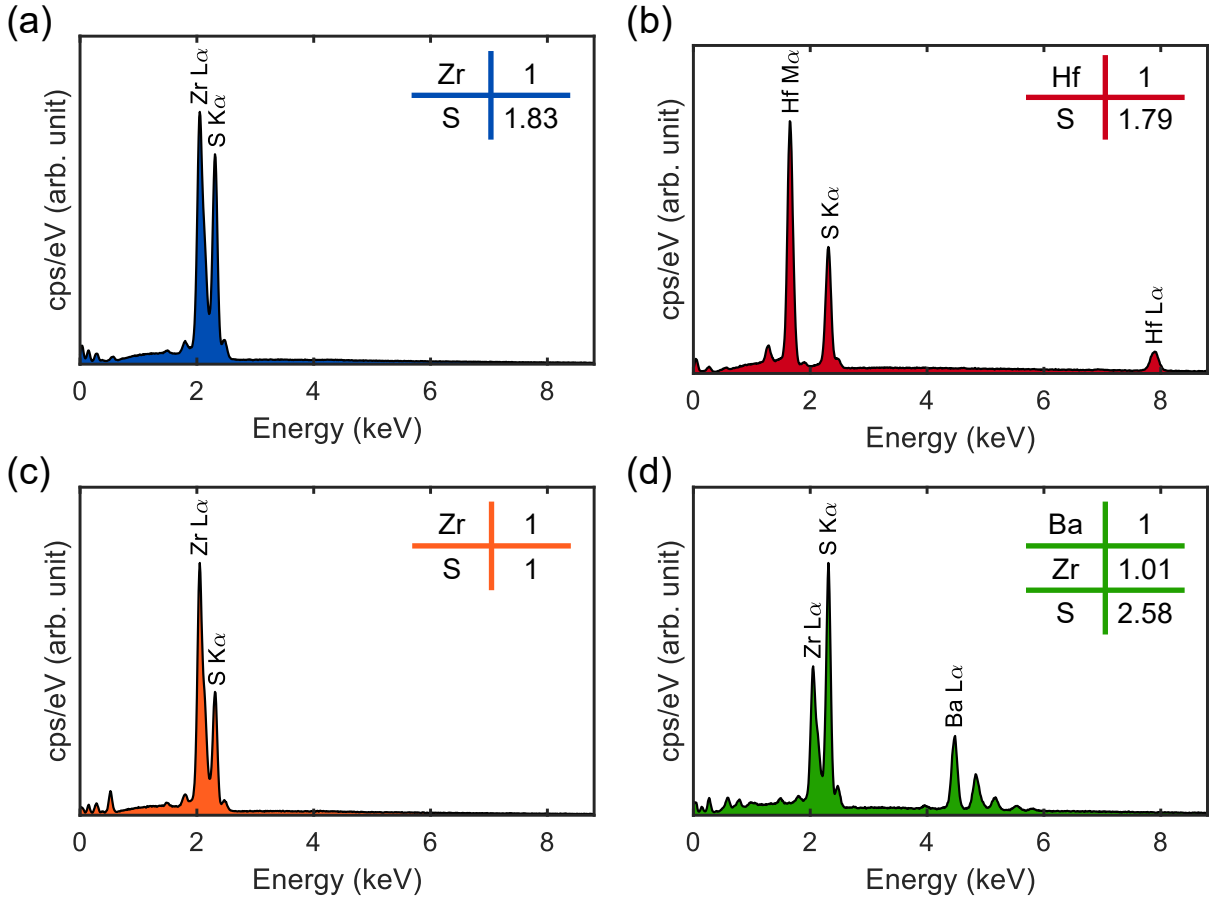

**Figure S2:** EDS spectra for **(a)** $ZrS_2$, **(b)** $HfS_2$, **(c)** ZrOS, and **(d)** $BaZrS_3$.

EDS spectra measured for the single crystals of $ZrS_2$, $HfS_2$, ZrOS, and $BaZrS_3$ are shown in **Figure S2**, demonstrating the stoichiometric ratio of the constituent elements. EDS maps were collected for all the synthesized elements to confirm a homogeneous distribution of corresponding elements. In the case of ZrOS, maps were collected for Zr and S. All the single crystals demonstrate a uniform distribution of constituting elements in a stoichiometric ratio.

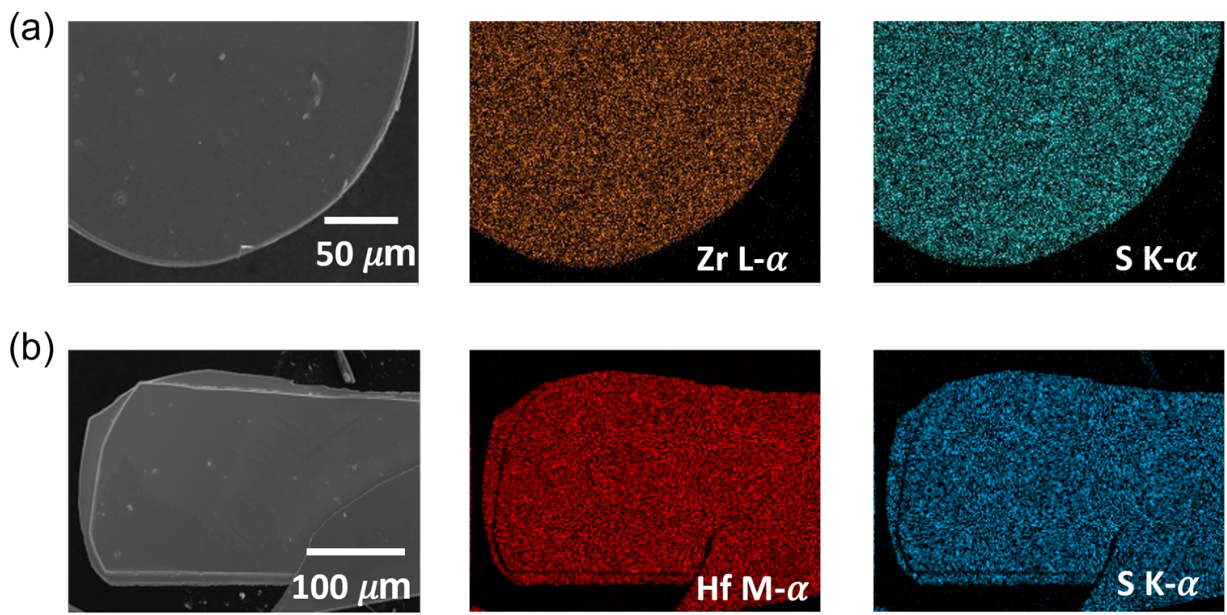

**Figure S3:** EDS Maps for **(a)** ZrS$_2$ and **(b)** HfS$_2$ plates.

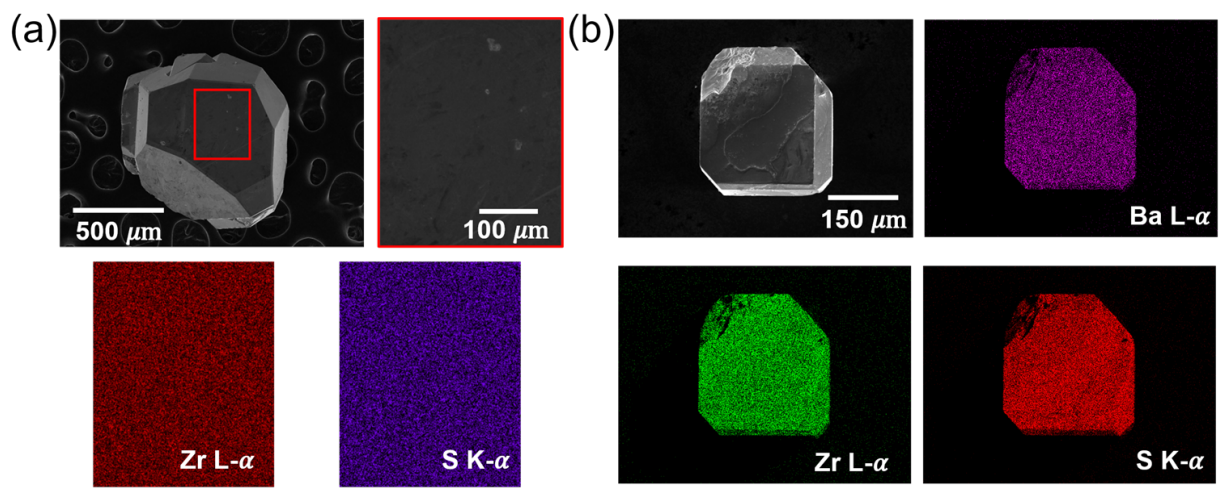

**Figure S4:** EDS Maps for **(a)** ZrOS and **(b)** BaZrS$_3$ single crystals.

## III. High-Resolution Thin Film Diffraction Measurements

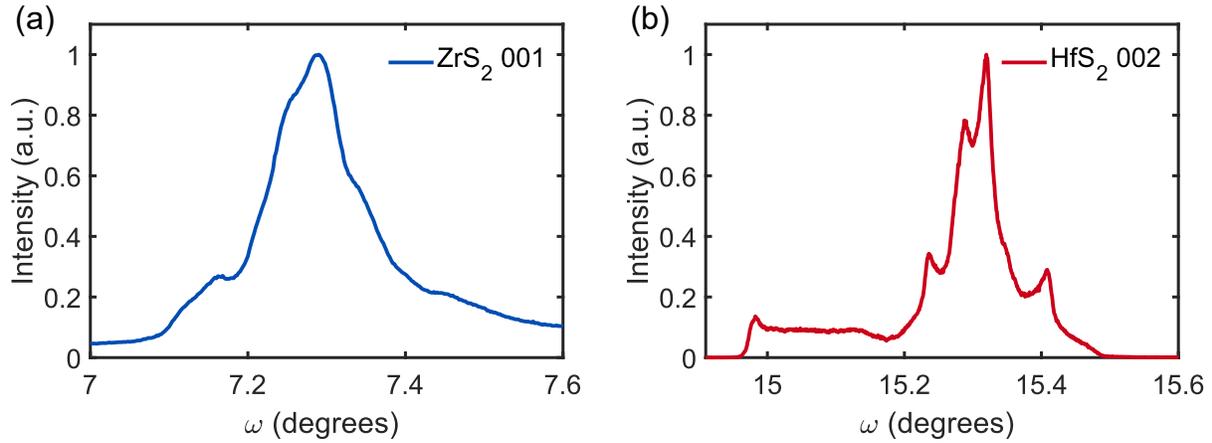

**Figure S5:** High resolution rocking curve scans for **(a)** 001 reflection of ZrS$_2$ and **(b)** 002 reflection of HfS$_2$.

The rocking curves collected for different 00$l$-type reflections for both ZrS$_2$ and HfS$_2$ demonstrate that the single crystals are twinned, as indicated by the presence of multiple peaks. In the case of BaZrS$_3$ and ZrOS, the narrow RC curves indicate the high quality of the single crystals. In the case of ZrOS, the facet orientations were also confirmed with SC-XRD measurements.

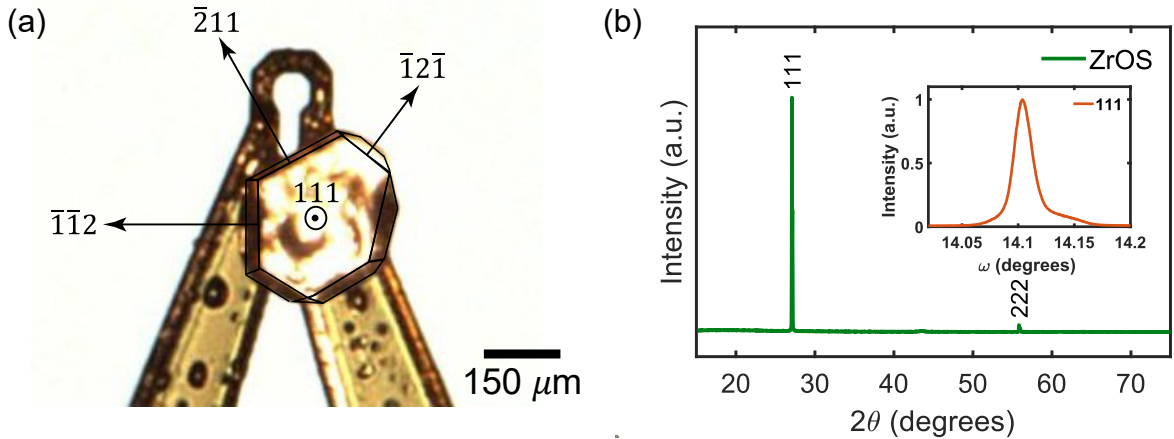

**Figure S6:** Orientation determination for ZrOS crystals. **(a)** ZrOS crystal mounted on a Kapton loop, with different facets labeled. **(b)** Out-of-plane X-ray diffraction pattern of ZrOS. The inset shows the rocking curve scan for the 111 reflection.

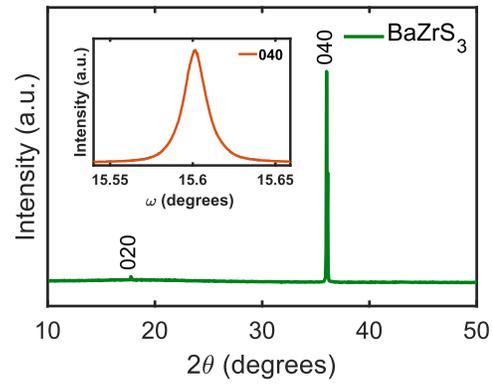

**Figure S7:** Out-of-plane X-ray diffraction pattern of BaZrS$_3$. The inset shows the rocking curve scan for the 040 reflection for BaZrS$_3$.

## IV. TRPL Measurements on ZrS₂ and HfS₂

Time-resolved photoluminescence measurements were carried out for ZrS$_2$ and HfS$_2$ at room temperature and 4K. Standard PL measurements were taken first using a 470 nm laser excitation. In the case of ZrS$_2$, since two peaks are observed, filters were used to scan the peaks individually, as shown in **Figure 1(d)** and **Figure S8(a)**. TRPL measurements were then taken for each peak using a repetition rate of 60 MHz at room temperature, and 20 MHz for 4 K. For room temperature TRPL curves, good fitting was observed for a single exponential model (**Figure 1(e, f)**), which was then used to extract the relaxation lifetime. The following equation was used:

$$y = A_1 exp\left(-\frac{t}{\tau_1}\right) + y_0$$

Relaxation lifetimes of 0.38 ns and 0.33 ns are observed for peaks A and B, respectively. For TRPL measurements at 4 K shown in **Figure S8(b, c)**, a bi-exponential model was used to fit the decay curves to extract the relaxation lifetimes using the following equation:

$$y = A_1 exp\left(-\frac{t}{\tau_1}\right) + A_2 exp\left(-\frac{t}{\tau_2}\right) + y_0$$

The relaxation lifetimes ($\tau_1$ and $\tau_2$) for peak A are 0.57 ns and 6.39 ns, with $A_1$ and $A_2$ being 95.57 % and 4.43 %, respectively. For peak B, relaxation lifetimes ($\tau_1$ and $\tau_2$) are 0.34 ns and 3.12 ns, with $A_1$ and $A_2$ being 99.2 % and 0.8%. The average lifetime was then calculated using $\tau_{av} = A_1\tau_1 + A_2\tau_2$. An average lifetime of ~0.82 ns is observed for peak A and ~0.36 ns for peak B.

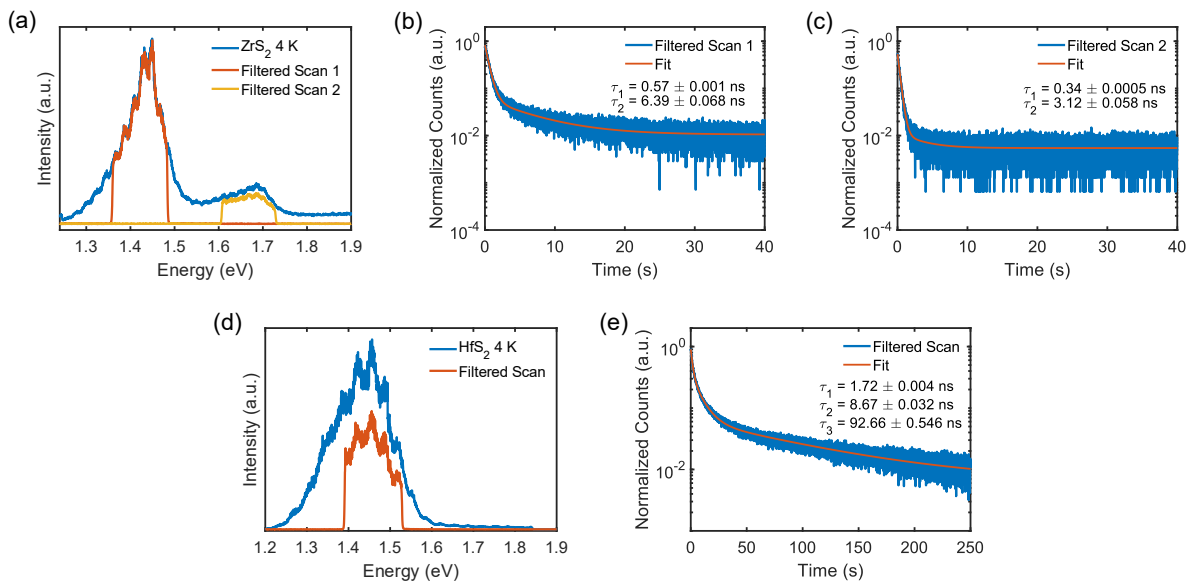

**Figure S8:** TRPL measurements for ZrS$_2$ and HfS$_2$ at 4K. **(a, d)** PL measurement for ZrS$_2$ and HfS$_2$ showing the filtered scans used for TRPL studies. TRPL measurements and fits for filtered scans for **(b, c)** ZrS$_2$ and **(e)** HfS$_2$, respectively.

In the case of HfS$_2$, TRPL measurements were performed using a repetition rate of 3 MHz at both room temperature and 4 K. At room temperature, different peaks observed in the standard PL measurement due to Fabry-Pérot interference fringes were used to get filtered scans, as shown in **Figure 1(g)**. TRPL measurements were carried out for these filtered scans. Different exponential decay models were tried to select the best fit, which was observed for the tri-exponential fit. The fit was carried out to extract the relaxation lifetimes, as shown in **Figure 1(h, i)**. The lifetimes observed for both peaks are similar: a fast relaxation lifetime of ~1 ns, an intermediate lifetime of ~5 ns, and a slow relaxation lifetime of 60-70 ns. Such a tri-exponential decay fit indicates multiple recombination pathways being active and might involve traps/ defect-assisted relaxation processes. The following equation was used to fit for tri-exponential decay:

$$y = A_1 exp\left(-\frac{t}{\tau_1}\right) + A_2 exp\left(-\frac{t}{\tau_2}\right) + A_3 exp\left(-\frac{t}{\tau_3}\right) + y0$$

The average lifetime was then calculated using $\tau_{av} = A_1\tau_1 + A_2\tau_2 + A_3\tau_3$, yielding ~0.93 ns and ~0.98 ns for filtered scans 1 and 2, respectively. The fit results have been tabulated in **Table S1**.

A filtered scan centered around the PL peak was used for TRPL measurements at 4 K, as shown in **Figure S8(d)**. The tri-exponential model was used to fit the TRPL decay curve and extract the relaxation lifetimes, as shown in **Figure S8(e)**. The lifetimes ($\tau_1$, $\tau_2$ and $\tau_3$) are 1.72 ns, 8.67 ns, and 92.66 ns, with $A_1$, $A_2$ and $A_3$ being 98.32 %, 1.54 % and 0.14 %, respectively. This gives an average relaxation lifetime of 1.96 ns for HfS$_2$ at 4 K.

**Table S1** includes the carrier lifetimes reported for bulk crystals of various TMDCs using TRPL results at room temperature. The shorter lifetimes reported for MoS$_2$, WS$_2$, and WSe$_2$[1] are 0.75 ns, 0.76 ns, and 1.6 ns, respectively, which are comparable to the single lifetime for ZrS$_2$ (0.3-0.4 ns), and the shortest lifetime observed for HfS$_2$ (~1 ns) crystals.

**Table S1: Room temperature TRPL lifetimes for TMDCs.**

| Material | $\tau_1$ (ns) | $A_1$(%) | $\tau_2$ (ns) | $A_2$(%) | $\tau_3$ (ns) | $A_3$(%) | $\tau_{av}$ (ns) |
|---|---|---|---|---|---|---|---|
| ZrS$_2$ FS1 | 0.38 | - | - | - | - | - | 0.38 |
| ZrS$_2$ FS2 | 0.33 | - | - | - | - | - | 0.33 |
| HfS$_2$ FS1 | 0.93 | 99.9893 | 5.07 | 0.0104 | 68.77 | 0.0003 | 0.93 |
| HfS$_2$ FS2 | 0.98 | 99.947 | 5.09 | 0.051 | 62.16 | 0.002 | 0.98 |
| MoS$_2$[1] | 0.749 | 61.61 | 6.088 | 38.39 | - | - | 2.798 |
| WS$_2$[1] | 0.764 | 62.09 | 7.318 | 37.91 | - | - | 3.249 |
| WSe$_2$[1] | 1.663 | 57.69 | 18.019 | 42.31 | - | - | 8.583 |

*FS1 and FS2 stand for filtered scan 1 and filtered scan 2, respectively.

## V. ZrOS XEOL Measurements

X-ray excited optical luminescence (XEOL) measurements were carried out using synchrotron X-ray exciting radiation of 9.67 keV energy for ZrOS single crystals. A clean peak around 2.48 eV is observed, confirming the PL peak observed for ZrOS using 375 nm pulsed laser excitation. Furthermore, XEOL mapping measurements were performed using the emission energy of 2.48 nm, as shown in **Figure S9(b)**, exhibiting different facets of the ZrOS single crystal.

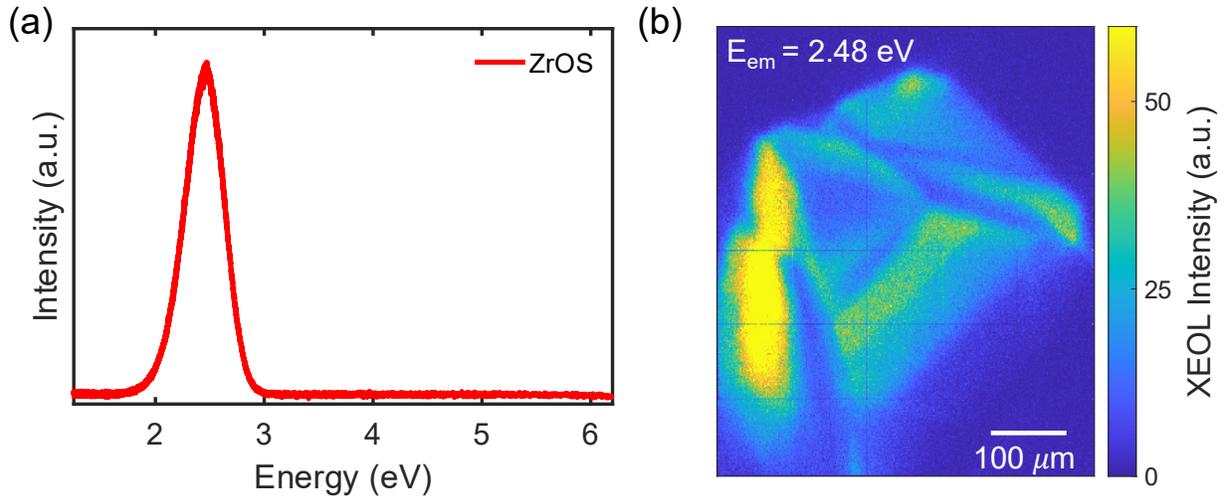

**Figure S9:** XEOL measurements on ZrOS single crystals. **(a)** XEOL spectrum exhibiting a peak around 2.48 eV, and **(b)** Emission map collected for ZrOS crystal at 2.48 eV, demonstrating different crystal facets.

## VI. SC-XRD Tables for ZrS$_2$, HfS$_2$, ZrOS and BaZrS$_3$

**Table S2. Data collection and refinement statistics of ZrS$_2$ at 290 K**

| | | |
|---|---|---|
| Diffractometer | \multicolumn{2}{c}{Rigaku XtaLAB Synergy-S} | |
| Source | PhotonJet (Mo) X-ray Source | |
| Wavelength | 0.71073 Å | |
| Temperature | 290 K | |
| Space group | $P\bar{3}m1$ | |
| **Cell dimensions** | | |
| $a, b, c$ (Å) | 3.66360(10), 3.66360(10), 5.8282(2) | |
| $\alpha, \beta, \gamma$ (°) | 90, 90, 120 | |
| Volume (Å$^3$) | 67.746(4) | |
| Density (g/cm$^3$) | 3.808 | |
| **Integration and Scaling** | | |
| Resolution (Å) | 5.83 – 0.65 | 5.83 – 0.70 |
| Completeness (%) | 93.8 | 94.2 |
| Redundancy | 17.9 | 20.8 |
| Mean $F^2 / \sigma(F^2)$ | 72.88 | 76.28 |
| $R_{int}$ | 0.051 | 0.050 |
| $R_{\sigma B}$ | 0.013 | 0.012 |
| **Statistics of Reflections** | | |
| $h_{min}, h_{max}$ | -5, 5 | |
| $k_{min}, k_{max}$ | -5, 5 | |
| $l_{min}, l_{max}$ | -8, 9 | |
| $R_\sigma, R_{int}$ | 0.0161, 0.0512 | |
| $\theta_{min}, \theta_{full}, \theta_{max}$ (°) | 3.496, 25.242, 33.282 | |
| Laue fraction max / full | 0.984 / 1.000 | |
| **Refinement** | | |
| Weighting | $w = 1/[\sigma^2(F_o^2) + (0.0264P)^2]$ where $P = (F_o^2 + 2F_c^2)/3$ | |
| No. reflections | 127 | |
| No. reflections [$I > 2\sigma(I)$] | 121 | |
| No. parameters | 8 | |
| No. constraints | 0 | |
| $R_1, wR_2$ [all data] | 0.0170, 0.0421 | |
| $R_1, wR_2$ [$I > 2\sigma(I)$] | 0.0183, 0.0442 | |
| GoF | 1.209 | |

**Table S3. Atomic coordinates and equivalent isotropic atomic displacement parameters for ZrS$_2$ at 290 K**

| Atom | x/a | y/b | z/c | U(eq) |
|---|---|---|---|---|
| Zr1 | 0 | 1.0000 | 1.0000 | 0.0115(2) |
| S1 | 0.333333 | 0.666667 | 0.75185(13) | 0.0100(2) |

**Table S4. Anisotropic atomic displacement parameters for ZrS$_2$ at 290 K**

| Atom | $U_{11}$ | $U_{22}$ | $U_{33}$ | $U_{23}$ | $U_{13}$ | $U_{12}$ |
|---|---|---|---|---|---|---|
| Zr1 | 0.0080(2) | 0.0080(2) | 0.0184(3) | 0 | 0 | 0.00402(12) |
| S1 | 0.0093(3) | 0.0093(3) | 0.0114(4) | 0 | 0 | 0.00464(13) |

**Table S5. Data collection and refinement statistics of HfS$_2$ at 290 K**

| Diffractometer | Rigaku XtaLAB Synergy-S | |
|---|---|---|
| Source | PhotonJet (Mo) X-ray Source | |
| Wavelength | 0.71073 Å | |
| Temperature | 290 K | |
| Space group | $P\bar{3}m1$ | |
| **Cell dimensions** | | |
| a, b, c (Å) | 3.6290(2), 3.6290(2), 5.8538(4) | |
| α, β, γ (°) | 90, 90, 120 | |
| Volume (Å$^3$) | 66.764(9) | |
| Density (g/cm$^3$) | 6.034 | |
| **Integration and Scaling** | | |
| Resolution (Å) | 5.83 – 0.65 | 5.83 – 0.70 |
| Completeness (%) | 96.1 | 100 |
| Redundancy | 17.5 | 19.8 |
| Mean $F^2/\sigma(F^2)$ | 45.97 | 48.60 |
| $R_{int}$ | 0.056 | 0.055 |
| $R_{\sigma B}$ | 0.028 | 0.024 |
| **Statistics of Reflections** | | |
| $h_{min}, h_{max}$ | -5, 5 | |
| $k_{min}, k_{max}$ | -5, 5 | |
| $l_{min}, l_{max}$ | -8, 8 | |
| $R_\sigma, R_{int}$ | 0.0286, 0.0553 | |
| $\theta_{min}, \theta_{full}, \theta_{max}$ (°) | 3.480, 25.242, 33.328 | |
| Laue fraction max / full | 0.961 / 1.000 | |
| **Refinement** | | |
| Weighting | $w = 1/[\sigma^2(F_o^2) + (0.0766P)^2]$ where $P = (F_o^2 + 2F_c^2)/3$ | |
| No. reflections | 124 | |

| | |
|---|---|
| No. reflections [$I > 2\sigma(I)$] | 124 |
| No. parameters | 7 |
| No. constraints | 0 |
| $R_1$, $wR_2$ [all data] | 0.0430, 0.1052 |
| $R_1$, $wR_2$ [$I > 2\sigma(I)$] | 0.0430, 0.1052 |
| GoF | 1.149 |

**Table S6. Atomic coordinates and equivalent isotropic atomic displacement parameters for HfS$_2$ at 290 K**

| Atom | x | y | z | U(eq) |
|---|---|---|---|---|
| Hf01 | 0 | 1.0000 | 0.5000 | 0.0138(5) |
| S002 | 0.333333 | 0.666667 | 0.2535(8) | 0.0124(8) |

**Table S7. Anisotropic atomic displacement parameters for HfS$_2$ at 290 K**

| Atom | $U_{11}$ | $U_{22}$ | $U_{33}$ | $U_{23}$ | $U_{13}$ | $U_{12}$ |
|---|---|---|---|---|---|---|
| Hf01 | 0.0092(5) | 0.0092(5) | 0.0230(7) | 0 | 0 | 0.0046(3) |
| S002 | 0.0100(12) | 0.0100(12) | 0.0173(17) | 0 | 0 | 0.0050(6) |

**Table S8. Data collection and refinement statistics of ZrOS at 295 K**

| | | |
|---|---|---|
| **Diffractometer** | Rigaku XtaLAB Synergy-S | |
| **Source** | PhotonJet (Mo) X-ray Source | |
| **Wavelength** | 0.71073 Å | |
| **Temperature** | 295 K | |
| **Space group** | $P2_13$ | |
| **Cell dimensions** | | |
| a, b, c (Å) | 5.701, 5.701, 5.701 | |
| α, β, γ (°) | 90, 90, 90 | |
| Volume (Å$^3$) | 185.3 | |
| Density (g/cm$^3$) | 4.993 | |
| **Integration and Scaling** | | |
| Resolution (Å) | 5.70 – 0.64 | 5.70 – 0.70 |
| Completeness (%) | 99.4 | 100 |
| Redundancy | 94.4 | 112 |
| Mean $F^2 / \sigma(F^2)$ | 116.41 | 121.01 |
| $R_{int}$ | 0.066 | 0.065 |
| $R_{\sigma B}$ | 0.012 | 0.008 |
| **Statistics of Reflections** | | |
| $h_{min}$, $h_{max}$ | -8, 8 | |
| $k_{min}$, $k_{max}$ | -8, 8 | |

| $l_{min}$, $l_{max}$ | -8, 8 |
|---|---|
| $R_\sigma$, $R_{int}$ | 0.0155, 0.0654 |
| $\theta_{min}$, $\theta_{full}$, $\theta_{max}$ (°) | 5.057, 25.242, 33.886 |
| Laue fraction max / full | 1.000 / 1.000 |
| Friedel fraction max / full | 1.000 / 1.000 |
| **Refinement** | |
| Weighting | $w = 1/[\sigma^2(F_o^2) + (0.0022P)^2 + 0.3184(P)]$ where $P = (F_o^2 + 2F_c^2)/3$ |
| No. reflections | 254 |
| No. reflections [$I > 2\sigma(I)$] | 245 |
| No. parameters | 11 |
| No. constraints | 0 |
| $R_1$, $wR_2$ [all data] | 0.0151, 0.0300 |
| $R_1$, $wR_2$ [$I > 2\sigma(I)$] | 0.0124, 0.0258 |
| GoF | 1.402 |

**Table S9.** Atomic coordinates and equivalent isotropic atomic displacement parameters for ZrOS at 295 K

| Atom | x/a | y/b | z/c | U(eq) |
|---|---|---|---|---|
| Zr01 | 0.92979(5) | 0.57021(5) | 0.42979(5) | 0.00524(18) |
| S002 | 0.66667(13) | 0.83333(13) | 0.16667(13) | 0.0062(3) |
| O003 | 0.8448(4) | 0.3448(4) | 0.1552(4) | 0.0059(7) |

**Table S10.** Anisotropic atomic displacement parameters for ZrOS at 295 K

| Atom | $U_{11}$ | $U_{22}$ | $U_{33}$ | $U_{23}$ | $U_{13}$ | $U_{12}$ |
|---|---|---|---|---|---|---|
| Zr01 | 0.00524(18) | 0.00524(18) | 0.00524(18) | -0.00021(10) | 0.00021(10) | -0.00021(10) |
| S002 | 0.0062(3) | 0.0062(3) | 0.0062(3) | -0.0002(3) | 0.0002(3) | -0.0002(3) |
| O003 | 0.0059(7) | 0.0059(7) | 0.0059(7) | -0.0003(8) | -0.0003(8) | 0.0003(8) |

**Table S11.** Data collection and refinement statistics of BaZrS$_3$ at 300 K

| Diffractometer | Rigaku XtaLAB Synergy-S |
|---|---|
| Source | PhotonJet (Mo) X-ray Source |
| Wavelength | 0.71073 Å |
| Temperature | 300 K |
| Space group | *Pnma* |
| **Cell dimensions** | |
| a, b, c (Å) | 7.0661(4), 9.9878(6), 7.0224(4) |

| α, β, γ (°) | 90, 90, 90 | |
|---|---|---|
| Volume (Å³) | 495.60(5) | |
| Density (g/cm³) | 4.352 | |
| **Integration and Scaling** | | |
| Resolution (Å) | 9.99 – 0.67 | 9.99 – 0.75 |
| Completeness (%) | 87.6 | 99.9 |
| Redundancy | 4.7 | 5.2 |
| Mean $F^2 / \sigma(F^2)$ | 23.98 | 25.24 |
| $R_{int}$ | 0.038 | 0.037 |
| $R_{\sigma B}$ | 0.028 | 0.026 |
| **Statistics of Reflections** | | |
| $h_{min}, h_{max}$ | -9, 10 | |
| $k_{min}, k_{max}$ | -14, 14 | |
| $l_{min}, l_{max}$ | -8, 10 | |
| $R_\sigma, R_{int}$ | 0.0373, 0.0324 | |
| $\theta_{min}, \theta_{full}, \theta_{max}$ (°) | 3.547, 25.242, 32.014 | |
| Laue fraction max / full | 0.873 / 1.000 | |
| **Refinement** | | |
| Weighting | $w = 1 / [\sigma^2(F_o^2) + (0.0091P)^2 + 2.1756(P)]$ where $P = (F_o^2 + 2F_c^2)/3$ | |
| No. reflections | 792 | |
| No. reflections [$I > 2\sigma(I)$] | 685 | |
| No. parameters | 30 | |
| No. constraints | 0 | |
| $R_1, wR_2$ [all data] | 0.0342, 0.0565 | |
| $R_1, wR_2$ [$I > 2\sigma(I)$] | 0.0280, 0.0551 | |
| GoF | 1.140 | |

**Table S12. Atomic coordinates and equivalent isotropic atomic displacement parameters for BaZrS$_3$ at 300 K**

| Atom | x/a | y/a | z/c | U(eq) |
|---|---|---|---|---|
| Ba01 | 0.53883(6) | 0.7500 | 0.00753(5) | 0.01676(14) |
| Zr02 | 0.5 | 0.5 | 0.5 | 0.00918(15) |
| S003 | 0.28719(14) | 0.53066(11) | 0.21308(14) | 0.0137(2) |
| S004 | 0.5041(2) | 0.25 | 0.4412(2) | 0.0156(3) |

**Table S13. Anisotropic atomic displacement parameters for BaZrS$_3$ at 300 K**

| Atom | $U_{11}$ | $U_{22}$ | $U_{33}$ | $U_{23}$ | $U_{13}$ | $U_{12}$ |
|---|---|---|---|---|---|---|
| Ba01 | 0.0178(2) | 0.0150(2) | 0.0174(2) | 0 | 0.00321(15) | 0 |
| Zr02 | 0.0084(3) | 0.0095(3) | 0.0097(3) | 0.00021(19) | 0.00008(19) | 0.00033(17) |

| Atom | $U_{11}$ | $U_{22}$ | $U_{33}$ | $U_{23}$ | $U_{13}$ | $U_{12}$ |
|---|---|---|---|---|---|---|
| S003 | 0.0111(4) | 0.0183(5) | 0.0118(4) | 0.0011(4) | -0.0047(4) | -0.0008(4) |
| S004 | 0.0214(8) | 0.0078(7) | 0.0178(7) | 0 | -0.0020(6) | 0 |

**SI References**